\documentclass[12pt,a4paper
]{article}
\usepackage[russian,english]{babel}
\usepackage[cp1251]{inputenc}
\usepackage[T2A]{fontenc}
\usepackage{latexsym}
\usepackage{amssymb}
\usepackage{amsmath}
\usepackage{graphicx}

\newcommand{\be}{\begin{equation}}
\newcommand{\ee}{\end{equation}}
\newcommand{\p}[1]{(\ref{#1})}
\def\a{\alpha}

\def\b{\beta}
\def\p{\partial}

\def\t{\theta}
\def\tr{{\rm tr}}
\def\tr{{\rm tr}\,}
\def\Tr{{\rm Tr}\,}
\def\cN{{\cal N}}
\def\cD{{\cal D}}

\def\bea{\begin{eqnarray}}
\def\eea{\end{eqnarray}}

\def\cN{{\cal N}}

\def\cF{{\cal F}}
\def\cH{{\cal H}}
\def\f{\frac}

\def\tr{{\rm tr}\,}

\def\d{\delta}
\def\q{\quad}
\def\g{\gamma}

\def\l{\label}
\def\ve{\varepsilon}

\def\sB{\stackrel{\frown}{\Box}}

\def\l{\ldots}
\def\cF{\cal F}
\def\cH{\cal H}
\def\cV{\cal V}
\def\cW{\cal W}
\def\cY{\cal Y}

\date{\it  }
\begin{document}

\begin{center}
\vspace{1cm} {\Large\bf Leading low-energy effective action in the
$6D$ hypermultiplet theory on a vector/tensor background
\vspace{1.2cm} }

\vspace{.2cm}
 {I.L. Buchbinder$^{a}$,
 N.G. Pletnev$^{b}$
}

\vskip 0.6cm { \em \vskip 0.08cm \vskip 0.08cm $^{a}$Department of
Theoretical Physics, Tomsk State Pedagogical University,\\ Tomsk,
634061 Russia; joseph@tspu.edu.ru\\
and\\ National Research Tomsk State University, Tomsk, 634050 Russia
\vskip 0.08cm \vskip 0.08cm
$^{b}$Department of Theoretical Physics, Sobolev Institute of Mathematics\\
and National Research Novosibirsk State University,\\
 Novosibirsk, 630090 Russia; pletnev@math.nsc.ru

 }
\vspace{.2cm}
\end{center}

\begin{abstract}
 %%%%%%%%%%%%%%%%%%%%%%%%%%%%%%%%%
We consider a six dimensional (1,0) hypermultiplet model coupled to
an external field of vector/tensor system and study the structure of
the low-energy effective action of this model. Manifestly a (1,0)
supersymmetric procedure of computing the effective action is
developed in the framework of the superfield proper-time technique.
The leading low-energy contribution to the effective action is
calculated.

%%%%%%%%%%%%%%%%%%%%%%%%%%%%%%%%%
\end{abstract}

\section{Introduction}

In our recent paper \cite{BP15} we have developed the harmonic
superfield formulation of the $6D$ vector/tensor system and
constructed its coupling to $6D$ hypermultiplet. One of the
important and interesting applications of such a coupling is a
problem of the effective action induced by the hypermultiplet
interaction with the vector/tensor background. In the paper
\cite{BP15} we introduced the corresponding effective action, which
is a harmonic superfield functional of the vector/tensor system, and
computed the structure of its divergences. The present paper is
devoted to continuation of the research originated in \cite{BP15}.
Our basic purpose here is to calculate the finite first leading
low-energy contribution to the effective action. The main motivation
to studying the low-energy effective action in the theory under
consideration is related to a description of the low-energy dynamics
of M5-branes in terms of field theory.

As it is known, the M2- and M5-branes arise as states of the strong
coupling phase of M-theory (see e.g. \cite{BLMP} for a review and
references).  The low-energy dynamics of a single M5-brane is
described by the Abelian $\cN = (2,0)$ tensor multiplet \cite{FS}.
The field content of this multiplet is determined as follows. There
are five scalars which arise as the Goldstone bosons from
spontaneous breaking of the eleven-dimensional translational
invariance by a brane. The M5-brane is a 1/2-BPS object and
therefore there are eight fermionic degrees of freedom. The three
additional bosonic degrees of freedom are provided by an Abelian
2-form gauge field $B_{ab}$ which has a self-dual field strength
$H_{abc}$. This 2-form originates from breaking the gauge symmetry
of the 3-form potential which exists in M-theory. However a
Lagrangian description of such a system faces a problem: the kinetic
term for the 2-form gauge field is identically zero because of the
self-duality condition. In the non-Abelian case, there is an
additional problem since an appropriate generalization of the tensor
gauge symmetry is still unknown \cite{bekaert}\footnote{Various
proposals for dealing with this problem have been suggested (see
e.g. \cite{BSS} for a review and references).}. In addition, there
are the inevitable problems of quantization of such models and
whether the conformal symmetry is preserved at the quantum level.

The low energy theory of multiple M5-branes is an interacting
six-dimensional conformal field model with (2,0) supersymmetry (see
e.g. \cite{Exact} for a review and references.). The existence of
such field theories, as well as all of their known properties, have
originated from string theory, where they occur in various related
contexts: the IR limit of the M5 or IIA NS-5 brane world-volume
theory, IIB string theory on a ALE singularity \cite{witten}, M
theory on $AdS_7 \times S^4$ \cite{mald}, etc. The IR-limit of these
theories are (2,0) superconformal field models which obey an
ADE-classification: $SU(N)$, $SO(2N)$, or $E_{6,7,8}$ \cite{ADE},
but have no other parameters. It is worth pointing out that all that
is known about an interacting $6D$, $\cN = (2,0)$ field theory has
been obtained from string theory. In particular, the non-trivial
$SO(5)_R$ ’t Hooft anomaly  was found in \cite{an} in the context of
11d M- theory, which gave the anomaly for the case $G = SU(N)$,
realized as N parallel M5 branes. The corresponding anomaly
coefficient for the $SU(N)$ case was found with help of M theory on
$AdS_7\times S^4$ in \cite{KT96} to be $c_{SU(N)} = N^3 - N$.

In a series of works \cite{SSWW} it was considered the possibility
of constructing the (2,0) theory of multiple M5-branes using (1,0)
supersymmetry in the framework  so-called the non-Abelian hierarchy
of $p$-form fields \cite{hierar}. In this case the following
supermultiplets are used: tensor multiplet, hypermultiplet and super
Yang-Mills multiplet. In the framework of these models the SYM
multiplet should be auxiliary analogous to non-propagating gauge
fields in the BLG or ABJM theory for multiple M2-branes. Such models
are parameterized by a set of dimensionless constant tensors, which
are constrained to satisfy a number of algebraic identities. A
concrete model is defined by the explicit choice of the gauge group
and representations and the above associated invariant tensors. All
these theories can be treated as belonging to the same universality
class of theories which are dual to $AdS_7\times S^4$ and possibly
describe multiple M5-branes. A several explicit examples which
satisfy all algebraic consistency conditions has been discussed in
the literature (see e.g. \cite{SSWW}).

Superfield formulation of the tensor hierarchy has been studied in
the paper \cite{IB13} where a set of constraints on the
super-$(p+1)$-form field strengths of non-Abelian super-$p$-form
potentials in the (1,0) $6D$ superspace has been proposed. In
\cite{BP15} we considered six dimensional hypermultiplet, vector and
tensor multiplet models in (1,0) harmonic superspace and discussed
the corresponding superfield actions (see also \cite{Z86},
\cite{ISZ}, \cite{Sok}). The superfield actions for a free (2,0)
tensor multiplet and for an interacting vector/tensor multiplet
system in terms of (1,0) superfields have been constructed for the
first time in \cite{BP15}. To construct (2,0) theory, one adds $n_T$
(1,0) superconformal hypermultiples to the above (1,0) vector/tensor
system. It is worth mentioning that there is no direct interaction
between hypermultiplets and tensor multiplets, a coupling between
these multiplets is provided by a vector multiplet (see e.g.
\cite{SSWW}). Such a coupling comes through the auxiliary fields,
which are described by the algebraic field equation
$d_{Irs}(Y^s_{ij}\phi^I-2\bar\lambda^s_{(i}\chi^I_{j)})-\l=0$. In
general, this equation implies constraints on the elementary fields
\cite{SSWW} but inclusion of Abelian factors or tensor multiplet
singlets, allows us to bypass constrains on the elementary fields
and, in particular, leads to the interaction terms of the form
${\cal L}_{\phi^0{\cF}^2}$. In that case there is a unique solution
for the auxiliary fields $Y_{ij}$. The resulting scalars can take
any values and then the vev of the tensor multiplet scalar acts as
an inverse Yang-Mills coupling constant in the conformal broken
phase. This effect is similar in many aspects of the ”M2 to D2”
scenario \cite{M2D2} proposed for the BLG theory which teaches us
that the M2-brane field theory is the strongly-coupled limit of the
D2-brane theory where the type IIA string theory transforms into
M-theory. Such a circumstance allows us to consider the Coulomb
brunch of the theory and study of the perturbative properties of the
models on this branch.

The next natural question is, what are the higher-order corrections
to the M5-brane action where the fields of the vector multiplet be
come dynamic degrees of freedom. One of the direct way to answer
this question is to derive the effective action by calculating the
open string scattering amplitudes. This program for the Abelian case
yielded the full higher-derivative purely bosonic terms in the
Dirac-Born-Infeld approximation \cite{BIaction}. In addition, there
exists a remarkable connection between ~i) partial supersymmetry
breaking, ~ii) nonlinear realizations of extended supersymmetry,
~iii) BPS solitons, and ~iv) nonlinear Born-Infeld-Nambu type
actions \cite{embed},  \cite{BIK}, \cite{ket98}\footnote{Due to the
large number of relevant papers we have no possibility to cite a
large number papers on these aspects.}. On the other hand, the
systems of D5-branes have complementary descriptions in terms of
gauge theory (see e.g. \cite{schwarzJ}). As one of the consequences,
the leading-order interaction potential between separated branes
admits representation as a leading term in the quantum gauge theory
effective action. The agreement between the supergravity and the
gauge theory expressions for the potential is possible because of
the existence of certain non-renormalization theorems on the gauge
theory side (see e.g. \cite{D4N4}). Since the hypermultiplet has a
universal coupling to the vector multiplet, one can expect that, in
the context of field theory, it will be possible to derive directly
the leading higher order 6D supersymmetric correction to the
classical action. Precisely this problem is considered in the
present paper.

We begin with harmonic superfield $6D$ hypermultiplet coupled to an
external field of vector/tensor system and compute the one-loop
effective action depending on the superfields of the vector/tensor
system. To develop the method of calculating of the effective action
and study of its possibilities we consider the simplest case when
all the fields are Abelian. As the result we find superfield action
which corresponds to the 6D (1,0) superconformal '$F^4$' term in the
components.

\section{Model of $6D$ hypermultiplet coupled to vector/tensor system}

We consider the hypermultiplet model coupled to an external field of
the vector/tensor system in the framework of the formalism of the
(1,0) harmonic superspace\footnote{We follow the harmonic superspace
conventions of \cite{gios} to which we refer for definitions,
notations and additional references. Its application to
vector/tensor system is discussed in \cite{BP15}.}. Our main aim is
to compute the leading low-energy contribution to the superfield
effective action depending on the superfields of the vector/tensor
system

Let us briefly discuss the structure of the vector/tensor system.
The (1,0) superconformal $6D$ field theory of the vector/tensor
system describes a hierarchy of non-Abelian scalar, vector and
tensor fields $\{\phi^I, A_a^r, Y^{ij\  r}, B_{ab}^I, C_{abc\ r},
C_{abcd\ A}\}$ and their supersymmetric partners which are labeled
by the indices $r=1,\l, n_V$ and $I=1,\l,n_T$ (see the details e.g.
in \cite{SSWW}). The non-Abelian field strengths of the vector and
two-form gauge potentials are given as \be\label{calF}
{\cF}^r_{ab}=\p_{[a}A^r_{b]}-f_{st}^{\ \ r}A^s_a A^t_b +h^r_I
B^I_{ab}, \ee
$${\cH}^I_{abc}= \f12\cD_{[a}B^I_{bc]}+d^I_{rs}A^r_{[a}\p_b A^s_{c]}-
\f13f_{pq}^{\ \ s}d^I_{rs}A^r_{[a}A_b^pA_{c]}^q+g^{Ir}C_{abc\ r}.$$
Here  $f_{[st]}^{\ \ r}$ are the structure constants, $d^I_{(rs)}$
are the $d$-symbols, defining the Chern-Simons couplings, and
$h^r_I, g^{Ir}$ are the covariantly constant tensors, defining the
general St\"uckelberg-type couplings among the forms of different
degrees. The existence of the non-degenerate Lorentz-type metric
$\eta_{IJ}$, such that $h^r_I=\eta_{IJ}g^{Jr}$,
$b_{Irs}=2\eta_{IJ}d^J_{rs}$, is also assumed. The covariant
derivatives are defined as $\cD_a=\p_a-A^r_a X_r$ with the gauge
generators $X_r$ acting on the different fields as follows:
$X_r\cdot \Lambda^s\equiv -(X_r)_t^s\Lambda^t~,$ $X_r\cdot
\Lambda^I\equiv -(X_r)_J^I\Lambda^J~.$ The covariance of the field
strengths (\ref{calF}) requires that the gauge group generators in
the various representations should have the form
$$(X_r)_s^t=-f_{rs}^{\ \ t}+g^t_I d^I_{rs}, \q (X_r)_I^J=2d^J_{rs}g^s_I-g^{Js}d_{Isr},$$
in terms of the invariant tensors parameterizing the system (see the
details in \cite{SSWW}). The field strengths (\ref{calF}) are
defined in such a way that they transform covariantly under the set
of non-Abelian gauge transformations \be\label{trA}\d
A_a=\cD_a\Lambda^r -h_I^r\Lambda^I_a,\ee
$$\d B_{ab}^I=\cD_{[a}\Lambda^I_{b]}-2d^I_{rs}(\Lambda^r{\cF}^s_{ab}-
\f12A^r_{[a}\d A^s_{b]})-g^{Ir}\Lambda_{ab\ r}.$$

The superspace realization of the tensor hierarchy was developed in
the paper \cite{IB13} in framework of the conventional 6D, (1,0)
superspace by means of study of the consistency conditions for the
generalized Bianchi identities. In \cite{BP15} we reformulated the
6D hypermultiplet, vector and tensor multiplet models in (1,0)
harmonic superspace and discussed the corresponding superfield
actions. Further, we will use the results of the works \cite{IB13},
\cite{BP15}. It is convenient to introduce the generalized
superfield strength \be\label{genstrength} {\cW}^{i\a\ r}=W^{i\a \
r}+g_I^r{\cal V}^{i\a \ I}, \ee where the $W^{i\a \ r}$ is the
superfield strength of the super Yang-Mill theory (defined in
\cite{Z86}, \cite{ISZ}) and ${\cal V}^{i\a \ I}$ is the
superpotential of the tensor multiplet (defined in \cite{Sok}), and
write the generalized Bianchi identities in its terms. Then one can
see that the conventional strength $F_{ab}$ of the vector multiplet
and the potential $B_{ab}$ of the tensor multiplet enter into
${\cW}^{i\a\ r}$ in the gauge covariant form ${\cF}^r_{ab}=F^r_{ab}
+ g^r_I B^I_{ab}.$ The other superfield strengths of the
vector/tensor multiplet are defined as \be {\cY}^{++
r}=\f14\cD^{+}_\a{\cW}^{+\a r}, \q
g^r_I\Phi^I=\f14(\cD^-_\a{\cW}^{+\a  r}-\cD^+_\a{\cW}^{-\a r}), \ee
$$ \Psi^{\pm I}_\a
=-\f{i}{2}\cD^\pm_\a\Phi^I, \q
g^r_I{\cH}^I_{abc}=\cD_{[a}{\cF}^r_{bc]}.$$ The  algebra of the
covariant derivatives $\cD^\pm_\a, \cD^{\pm\pm}, \cD_{a}$ is
described in \cite{BP15}. By applying a harmonic-dependent gauge transformation, one can choose a $\lambda$-frame where $\cD^+_\a \rightarrow D^+_\a$, $\cD^{++}=D^{++} +{V}^{++}$, $\cD^{--}=D^{--} +{\cV}^{--}$,  with $V^{++}$  is the analytic prepotential for the off-shell vector multiplet, and the other harmonic connection ${\cV}^{--}$ is the linear combination of the non-analytic potential $V^{--}$ for vector multiplet and the potential ${\cV}^{(-2)}$ for on-shell tensor multiplet (see \cite{ISZ}, \cite{Sok}, \cite{BP15} for more details). By using these superfields one can define
the superfield action in harmonic superspace as follows \be\label{S}
S = \f18\int d\zeta^{(-4)} du \ g_{I r}\{\Phi^I \cD^{++}{\cal Y}^{++\
r}+D^+_\a\Phi^I \cD^{++}{\cal W}^{+{\alpha}r}\}~, \ee
where $d\zeta^{(-4)}$ denotes the analytic subspace integration measure. The action
(\ref{S}) depends both on superfields $V^{++}$, ${W}^{\a i}$ of the
vector multiplet and on superfields $\Phi$, ${\cV}^{\a i}$
responsible for the tensor multiplet. If a vev of $\Phi$ is a
constant $1/f^2$, this action takes the form  of SYM action
\cite{Z86}, \cite{ISZ}
$$S\sim \f{1}{f^2}\int d^6x d^8\t du V^{++}V^{--},$$
as discussed above. The equation of motion for this action is
$Y^{++}=(D^+)^4V^{--}=0~.$

As a further step towards to a (2,0) theory  it was proposed in the
papers \cite{SSWW} to complement the non-Abelian vector/tensor by
superconformal hypermultiplets and construct the corresponding
coupling. The Lagrangian for these theories consists of two pars.
One part involves vector and tensor multiplets, and the second part
contains hypermultiplets coupled to the vector/tensor system. These
two parts are independently (1,0) supersymmetric.

A conformally invariant hypermultiplet model can be formulated in
six dimensional (1,0) harmonic superspace \cite{ISZ}. The
corresponding superfield action in general case is written as
follows \be\label{hyper} S=-\f12\int d\zeta^{(-4)}du
(q^{+A}\cD^{++}q^+_A+ L^{(+4)}(q^+,u))~. \ee  The potential $L^{(+4)}(q^+,u)$
determines a hypermultiplet self-interaction \cite{gios}, it is irrelevant for
our purposes and will be omitted further. We want to emphasize that
the superfield $V^{++}$ here is related to the superfield ${\cal
V}^{--}$ through zero curvature equation (see \cite{gios}). The
superfield strengths ${\cW}^{+\a}=-\f14(D^+)^{3\a}{\cV}^{--}$,
involving the superfield ${\cV}^{--\ \ r}={V}^{--\ \
r}+g^r_I{\cV}_{T}^{--\ \ I}$, obey the Bianchi identities which
contain the superfields $\Phi$, $\Psi^i_\a$, ${\cH}_{abc}$ related
to tensor multiplet (see \cite{BP15} for the details). As a result
the action (\ref{hyper}) describes the interaction of a
hypermultiplet with a vector/tensor system.

\section{Construction of effective action}
We will discuss here the procedure of calculating the effective
action corresponding to the hypermultiplet theory in an external
field of a vector/tensor system (\ref{hyper}). The effective action
is defined by integrating out hypermultiplet and keeping the $U(1)$
vector/tensor system as a background.

A formal relation for the effective action follows from
(\ref{hyper}) in the form
\be\label{Tr}
\Gamma=i\Tr\ln\cD^{++}=-i\Tr\ln G^{(1,1)}~,
\ee
where the
$G^{(1,1)}(\zeta_1,\zeta_2)$ is the hypermultiplet Green function,
satisfying the equation:
$$\cD^{++}_1 G^{(1,1)}(\zeta_1,\zeta_2)=\d_A^{(3,1)}(\zeta_1,\zeta_2)~,$$
\be\label{GREEN}
G^{(1,1)}(1|2)=-\f{1}{4\sB_1}(\cD^+_1)^4(\cD^+_2)^4\d^{14}(z_1-z_2)\f{1}{(u^+_1u^+_2)^3}~.\ee
Here $\d^{14}(z_1-z_2)=\d^{6}(x_1-x_2)\d^{8}(\t_1-\t_2)$ is the
delta-function in conventional superspace,
$\d^{(3,1)}_A(\zeta_1,\zeta_2)$ is the appropriate covariantly
analytic delta-function
$\d^{(3,1)}_A(\zeta_1,\zeta_2)=(\cD^+)^4\d^{14}(z_1-z_2)\d^{(-1,1)}(u_1,u_2)$
and $(u^+_1u^+_2)^{-3}$ a special harmonic distribution \cite{gios}.
In eg. (\ref{GREEN}), ${\sB}$ is the covariantly analytic
d'Alembertian which arises when $(\cD^+)^4(\cD^{--})^2$  acts on the
analytical superfield \be\label{sBox}
\sB=-\f18(\cD^+)^4(\cD^{--})^2=\cD_a\cD^a+{\cW}^{+\a}\cD^-_\a
+{\cY}^{++}\cD^{--}-{\cY}^{+-}-\Phi ~. \ee  The operator ${\sB}$ (\ref{sBox})
possesses the important properties \be[\cD^+_\a,{\sB}]=0~,\ee
$$[\cD^{++},\sB]V^{(p)}={\cY}^{++}(p-1)V^{(p)}~.$$
where $V^{(p)}(\zeta, u)$ is an arbitrary analytic superfield of
$U(1)$ charge $p$. To prove the above identities, one should make
use of the following properties of the $6D$, (1,0) gauge covariant
derivatives in harmonic superspace \cite{Z86}, \cite{ISZ},
\cite{BP15} \be [\cD^+_\a,\cD^-_\b]=2i\cD_{\a\b}, \q
[\cD^\pm_\g,\cD_{\a\b}]=-2i\ve_{\a\b\g\d}{\cW}^{\pm\g}~.\ee The
field strength ${\cW}^{\pm\a}$ obeys the generalized vector/tensor
Bianchi identities \be\label{DW}\cD^-_\a {\cW}^{+\b\ r}=\d_\a^\b
({\cY}^{+-\ r}+\f12\Phi^Ig_I^r)+\f12 {\cF}_\a^{\b\ r}~,\q \cD_\a^\pm
{\cY}^{+-\ r}=\pm i(\cD_{\a\b}{\cW}^{\pm\b\ r}+i\cD^{\pm }_\a
\Phi^Ig_I^r )~.\ee These properties follow from the 6D (1,0)
vector/tensor multiplet formulation \cite{IB13} in conventional
superspace.

The  definition (\ref{Tr}) of the one-loop effective action is
purely formal. The actual evaluation of the effective action can be
done in various ways (see e.g. \cite{BUCH}, \cite{KUZ}). Further we
mainly will follow \cite{KUZ} with some special differences and use
the relation \be\label{Gamma}\Gamma=\Gamma_{y=0}+\int_0^1
dy\p_y\Gamma(y V)=-i\Tr\int_0^1 dy(V^{++}G^{(1,1)}(y)),\ee  where
\be \Tr(V^{++}G^{(1,1)})=\int du_1
d\zeta_{1}^{(-4)}V^{++}(1)G^{(1,1)}(1|2)|_{1=2}.~\ee Here
$G^{(1,1)}(yV)$ means the Green function depending on the superfield
$yV^{++}$. Now one substitutes the expression (\ref{GREEN}) for
Green function $G^{(1,1)}(1|2)$ into (\ref{Gamma}) and uses a
proper-time representation for the inverse operator
$\f{1}{\stackrel{\frown}{\Box}}$. To avoid the divergences in the
intermediate steps of calculation one considers the regularized
inverse operator in the form (${\omega}$ - regularization)
\be\label{inverse}-\f{1}{\stackrel{\frown}{\Box}}=\int_0^\infty
d(is) (is{\mu}^2)^{\omega}e^{is\stackrel{\frown}{\Box}-\ve s}. \ee
%Since we calculate here only the finite contributions to effective
%action we can put the regularization parameter ${\omega}=0$.
The divergent part of the effective action has already been found in
\cite{BP15}, it was shown that it defines a charge renormalization
in the vector/tensor action (\ref{S}), and a higher derivative SYM
action, found in \cite{ISZ}. We calculate the effective action in
the local approximation where the effective action is represented as
a series in background fields and their derivatives and expressed in
terms of the effective Lagrangian in the form \be\label{eff.lagr}
\Gamma = \int d\zeta^{(-4)}du {\cal L}^{(+4)}~.\ee The further
analysis is based on the following identity involving the product of
${\cal D}$-factors presenting in the Green function (see derivation
of this identity for $4D$ and $5D$ cases in \cite{KUZ})
\be\label{Delta}(\cD^+_1)^4(\cD^+_2)^4\f{1}{(u^+_1u^+_2)^3}=(\cD^+_1)^4
\{(u^+_1u^+_2)(\cD^-_1)^4 -(u^-_1u^+_2)\Delta^{--}
-4\sB\f{(u^-_1u^+_2)^2}{(u^+_1u^+_2)}\}~.\ee Here \be\label{DELTA}
\Delta^{--}=i\cD^{\a\b}\cD^-_\a\cD^-_\b+4{\cW}^{-\a}\cD^-_\a-(D^-_\a
{\cW}^{-\a})~.\ee

Now we will discusses the restrictions on background. To find the
leading low-energy contribution to effective action it is sufficient
to consider a covariantly constant vector/tensor multiplet in the
absence of the auxiliary fields ('on-shell' background)
\be\label{backgr} \cD_a{\cW}^{\pm \a}=0,\q {\cY}^{ij}=0~.\ee %To%provide consistency
For self-consistency of the relations (\ref{backgr}) we should
supplement the above relations by the following relations
\be\label{Phi} D^i_\a\Phi=0, \q\q D^i_\a {\cF}_{ab}=0~. \ee In this
case the operators ${\stackrel{\frown}{\Box}}$ and ${\Delta^{--}}$
take a simple form and depends only on the background fields ${\cal
W}^{+\alpha}$, $D^-_\a{\cal W}^{+\b}$ and ${\Phi}$. Since the form
of the effective Lagrangian is defined by the coefficients of these
operators we can conclude that on the background under
consideration, the effective Lagrangian should have the following
general form \be\label{eff.lagr_1} {\cal L}^{(+4)}= {\cal
L}^{(+4)}({\cal W}^{+\alpha}, D^-_\a{\cal W}^{+\b},\Phi)~.\ee
Further we will see that in leading approximation the effective
Lagrangian does not depend on $D^-_\a{\cal W}^{+\b}$.

\section{Leading low-energy contribution to effective action}
We will consider now a computation of  the leading low-energy
quantum contribution to the effective action. First of all we
substitute the expression for the Green function (\ref{GREEN}) into
the expression for effective action (\ref{Gamma}) and use the
identity (\ref{Delta}). It leads to \be\label{Gamma-1} \Gamma
=\frac{i}{4}\int_0^1 dy\int
d\zeta_{1}^{(-4)}duV_1^{++}\frac{1}{\sB_1}(\cD^+_1)^4
\{(u^+_1u^+_2)(\cD^-_1)^4 -(u^-_1u^+_2)\Delta_1^{--}\ee
$$-4\sB\f{(u^-_1u^+_2)^2}{(u^+_1u^+_2)}\}\d^{14}(z_1-z_2)|_{2=1}~. $$

To get the leading low-energy contribution to the effective action
one analyses the terms in the expression (\ref{Gamma-1}) for the
background under consideration. First of all we take into account
that we should use eight $D^\pm_\a$-factors in this expression to
eliminate the $\delta$-function of anticommuting variables via the
identity \be (D^+)^4(D^-)^4\d^8(\t-\t')|_{\t=\t'}=1~.\ee

Consider the last term in (\ref{Gamma-1}). We see that the operator
$\sB$ is cancelled and then there is no enough number of $D$-factors
to eliminate the above $\delta$-function. Therefore this term is
zero. Now consider the first term in (\ref{Gamma-1}). This term was
analyzed in \cite{BP15}, it was shown that it is proportional to
${\cY}^{++}$, which is equal to zero on the background under
consideration.\footnote{This terms determines the divergences of the
effective action \cite{BP15}. In particular, it means that the
effective action is finite on the background under consideration.}
Now let us analyse the contributions of ${\Delta^{--}}$. the third
term here is proportional to ${\cY}^{--}$ and hence, it vanishes on
the background under consideration. Now we will use the proper-time
representation (\ref{inverse}) of the inverse operator $\sB$ in
(\ref{Gamma-1}) and expend
$e^{is\stackrel{\frown}{\Box}}=e^{is(\Box-\Phi)}e^{is{\cW}^+\cD^-}$
in the power series in ${\cal W}^{+\a}{\cal D}^{-}_{\a}$. The
leading contribution arises in the third order in this expansion.
Consider the contribution of the second term in (\ref{DELTA}) after
the above expansion. Schematically it has the form
$$\int d\zeta^{(-4)}du V^{++}{\cW}^-({\cW}^+)^3=\int
d\zeta^{(-4)}du V^{++}\cD^{--}({\cW}^+)^4 %= -\int
%d\zeta^{(-4)}du\cD^{--}V^{++}({\cW}^+)^4$$$$= -\int
%d\zeta^{(-4)}du\cD^{++}{\cV}^{--}({\cW}^+)^4
=\int
d\zeta^{(-4)}du{\cV}^{--}\cD^{++}({\cW}^+)^4=0.$$

Thus, the leading low-energy contribution to effective action is
given by the following expression \be\label{Gamma-2}\Gamma=-
\f{i}{4}\int_0^1 dy \int_0^\infty d(is)\int d\zeta^{(-4)}duV^{++}i
\cD^{\a\b}\cD^-_\b\cD^-_\a \f16(is{\cW}^{+\d}\cD^-_\d)^3
e^{is(\Box-\Phi)}(D^+)^4\delta^{14}(z-z')|_{2=1}~.\ee Then let us
integrate by parts with respect of the operator
$\cD^{\a\b}\cD^-_\b$. The following transformations of $\int
d\zeta^{(-4)}du V^{++}i\cD^{\a\b}\cD^-_\b {\cal L}^{(+3)}_\a$ look
schematically like
%$$\int d\zeta^{(-4)}du V^{++}i\cD^{\a\b}\cD^-_\b {\cal L}^{(+3)}_\a%=\int \cD^-_\b i\cD^{\a\b}V^{++} {\cal L}^{(+3)}_\a
$$=\int d\zeta^{(-4)}du \cD^-_\b \f14\ve^{\a\b\g\d}D^+_\g \cD^-_\d
V^{++} {\cal L}^{(+3)}_\a= \int d\zeta^{(-4)}du \cD^-_\b
\f14\ve^{\a\b\g\d}D^+_\g (\cD^{--}D^+_\d) V^{++} {\cal
L}^{(+3)}_\a$$$$ =-\int d\zeta^{(-4)}du \cD^-_\b
\f14\ve^{\a\b\g\d}D^+_\g D^+_\d (\cD^{--}V^{++}) {\cal L}^{(+3)}_\a=
-\int d\zeta^{(-4)}du \cD^-_\b \f14\ve^{\a\b\g\d}D^+_\g D^+_\d
(\cD^{++}{\cV}^{--}) {\cal L}^{(+3)}_\a$$$$ =\int d\zeta^{(-4)}du
D^+_\b \f14\ve^{\a\b\g\d}D^+_\g D^+_\d {\cV}^{--} {\cal
L}^{(+3)}_\a=\int d\zeta^{(-4)}du d\zeta^{(-4)}du{\cW}^{+\a}{\cal
L}^{(+3)}_\a.$$ The expression ${\cal L}^{(+3)}_\a$ is seen from
(\ref{Gamma-2}). Here we have used that $D^+_\a V^{++}=0,$ $D^{++}
(W^+)^3=0$ and that for a non-zero result there should be eight
$D$-factors acting on ${\delta}^{8}(\theta-\theta')$. As a result we
obtain that the effective Lagrangian depends only on ${\cal
W}^{+\a}$ and ${\Phi}$. After these transformations one gets the
integrand in (\ref{Gamma-2}) in the form
$$\sim (is)^3{\cW}^{+\a}{\cW}^{+\b}{\cW}^{+\g}{\cW}^{+\d}\cD^-_\a\cD^-_\b\cD^-_\g\cD^-_\d (\cD^+)^4\delta^{8}(\t-\t')
\sim (is)^3({\cW}^+)^4~.$$ By using the relation
$e^{is\Box}\d^6(x-x')|_{x=x'}=\f{i}{(4\pi is)^3} $, we finally
obtain
\be\label{W4}\Gamma[{\cW}^+, {\Phi}]=\f{1}{12(4\pi)^3}\int
d\zeta^{(-4)}du\f{({\cW}^+)^4}{\Phi}~.\ee
The effective action is
given as an integral over the analytic subspace of harmonic
superspace of the effective Lagrangian ${\cal L}^{(+4)}.$ It is
necessary to point out here that this effective Lagrangian satisfies
the condition of analyticity only on the background under
consideration where ${\cal D}^{+}_{\a}{\cal W}^{+\a}=0$ and
${\Phi}=\mbox{const}.$ For a generic background we should take into
account the terms containing the superfields ${\cal Y}^{++}$ and the
derivatives of the superfields ${\cW}, \Phi$, but all this lies
beyond the leading low-energy approximation.

Now we will consider the component structure of the effective
Lagrangian in the bosonic sector. By integrating over the
anticommuting coordinates $\int d^4\t^+=(\cD^-)^4$, one gets \be
(\cD^-)^4({\cW}^+)^4=\f{1}{4!}\ve_{\a\b\g\d}\ve^{\a'\b'\g'\d'}\cD^-_{\a'}
{\cW}^{+\a}\cD^-_{\b'} {\cW}^{+\b}\cD^-_{\g'} {\cW}^{+\g}\cD^-_{\d'}
{\cW}^{+\d}\ee
$$\sim
\f{1}{4!}\ve_{\a\b\g\d}\ve^{\a'\b'\g'\d'}\cN_{\a'}^\a \cN_{\b'}^\b
\cN_{\g'}^\g \cN_{\d'}^\d=\f{1}{4!}\det\cN~,$$ where we have denoted
$ \cN_\a^\b \equiv \cD^-_\a {\cW}^{+\b}|_{\t=0}~$ for (\ref{DW}). A
direct calculation of the determinant gives  \be\label{det}
\det\cN=(\cN)^4-6(\cN)^2\cN_\a^\b \cN_\b^\a+8(\cN)\cN_\a^\b
\cN_\b^\g \cN_\g^\a- 6\cN_\a^\b \cN_\b^\g \cN_\g^\d
\cN_\d^\a+3(\cN_\a^\b \cN_\b^\a)^2~,\ee where $(\cN)\equiv
\cN_\a^\a=2\Phi .$ It is also evident that $\tr\cN^3=0$. This
expression in the limiting case ${\Phi}=0$ is in agreement  with
earlier perturbative calculations of the low-energy effective action
of superstrings (see a review and references in \cite{BIaction} and
restrictions implied by supersymmetry in $6D$ \cite{ket98}). In the bosonic sector we have $\cN=\frac{1}{2}({\phi}+{\cal F})$ where
$${\cal F}_\a^{\ \b}=F_\a^{\ \b}+B_\a^{\ \b}.$$
It follows from the definition (\ref{calF}) in the Abelian case.
Here ${\phi}$ is a scalar
bosonic component of the superfield ${\Phi}$, $F_\a^{\
\b}=(\g^{ab})_\a^\b F_{ab}$ is the strength of Abelian vector field
and $B_\a^{\ \b}=(\g^{ab})_\a^\b B_{ab}$ is the antisymmetric tensor
field. Then it is evident that if we substitute relation (\ref{det})
into expression (\ref{W4}) and consider the bosonic sector, we get
the following terms $\phi^3$, $\phi {\cF}^2$, $\f{1}{\phi}{\cF}^4$
as the quantum corrections induced by the one-loop effect of the
hypermultiplet.

\section{Conclusion}
Let us briefly summarize the main results. We have considered a
problem of the induced effective action in the $6D$ (1,0)
hypermultiplet theory coupled to an external field of vector/tensor
system. The theory is formulated in six dimensional (1,0) harmonic
superspace in terms of an unconstrained analytic hypermultiplet
superfield in the external superfields corresponding to an Abelian
vector/tensor system. The effective action is formulated in the
framework of superfield proper-time technique which allows us to
preserve a manifest (1,0) supersymmetry. To calculate the low-energy
effective action it is sufficient to consider a special background
(\ref{backgr}), (\ref{Phi}). We have developed a generic procedure
for calculating the effective action on such a background and found
the leading low-energy contribution to the effective action
(\ref{W4}). The divergences in this theory have been computed in our
previous paper \cite{BP15}. It is worth mentioning that the
divergences are absent on the background (\ref{backgr}),
(\ref{Phi}).

We expect that the obtained results can have a relation to the
problem of the effective action of a single isolated D5-brane
\cite{schwarzJ}. However, to calculate the complete effective action
for such a D5-brane we should study a quantum vector/tensor +
hypermultiplet system. Of course, such a problem requires a special
consideration. Another aspect, which is essential for finding the
effective action of a D5-brane, is a necessity to curry out the
calculations on a conformally broken phase of the $6D$ non-Abelian
supersymmetric gauge theory (see definition of this phase e.g. in
\cite{SSWW}). Nevertheless, we hope that the methods, developed in
this paper, can be used to analyze the general problem of the
effective action of a D5-brane. The methods and results of the
present work can be generalized in the following directions: (i)
calculation of the low-energy effective action beyond the leading
approximation, (ii) calculation of the effective action in a
non-Abelian theory in the broken phase, (iii) calculation of the
effective action of the quantum vector/tensor+hypermultiplet system.

\section*{Acknowledgments }
%%%%%%%%%%%%%%%%%%%%%%%%%%%%%%%%%%%%%%%
The authors are thankful to E.
Buchbinder for useful discussions.
Work of I.L.B was supported by Ministry of Education and Science of
Russian Federation, project No 2014/387/122. N.G.P is grateful to
the RFBR grant, project No 15-02-03594 and LRSS grant, project No
88.2014.2 for partial support.

%%%%%%%%%%%%%%%%%%%%%%%%%%%%%%%%%%%%%%%%%%%%%%%
\bigskip
%%%%%%%%%%%%%%%%%%%%%%%%%%%%%%%


\begin{thebibliography}{000}
\baselineskip=14pt
\parskip=0.pt
\bibitem{BP15}I.L. Buchbinder, N.G. Pletnev,  Nucl.Phys. B 892 (2015) 21-48.
\bibitem{BLMP}J. Bagger, N. Lambert, S. Mukhi, C. Papageorgakis, Phys.Rept. 527 (2013) 1-100;
N. Lambert, Ann. Rev. Nucl. Part. Sci. 62 (2012) 285-313.
\bibitem{FS}S. Ferrara,  E. Sokatchev, Lett.Math.Phys. 51 (2000) 55-69;  J.Math.Phys. 42 (2001)
3015-3026.
\bibitem{bekaert}X. Bekaert, M. Henneaux and A. Sevrin,  Phys. Lett. B 468 (1999) 228;
Commun. Math. Phys. 224 (2001) 683.
\bibitem{BSS}I. Bandos, H. Samtleben, D. Sorokin,  Phys.Rev. D88 (2013) 2, 025024;
C-S. Chu,  Nucl.Phys. B866 (2013) 43-57.
\bibitem{Exact}J. Teschner, {\it Exact results on N=2 supersymmetric gauge theories}, e-Print: arXiv:1412.7145
[hep-th].
\bibitem{witten} E. Witten, {\it Some comments on string dynamics}, In *Los Angeles 1995, Future perspectives in string theory* 501-523,
e-Print: hep-th/9507121;
{\it Conformal Field
Theory In Four And Six Dimensions}, e-Print: arXiv:0712.0157
[math.RT]; N. Seiberg and E. Witten, Nucl. Phys. B 471, 121 (1996);
N. Seiberg, Phys. Lett. B 390, 169 (1997).
\bibitem{mald}J. M. Maldacena,  Adv. Theor. Math. Phys. 2 (1998) 231; O. Aharony, S.S. Gubser, J.M. Maldacena, H.
Ooguri, Y. Oz, Phys.Rept. 323 (2000) 183-386.
\bibitem{ADE}J. J. Heckman, D. R. Morrison and C. Vafa,  JHEP 1405, 028 (2014).
\bibitem{an}J. Harvey, R. Minasian, and G. Moore,  JHEP 9809 (1998) 004; K. A. Intriligator,
  Nucl.Phys. B581 (2000) 257-273;   JHEP 1410 (2014) 162.
\bibitem{KT96}I.R. Klebanov and A. Tseytlin,  Nucl. Phys. B 475 (1996) 164; M. Henningson and K. Skenderis,  JHEP 9807 (1998)
023; A.A. Tseytlin and K. Zarembo, Phys.Lett. B474 (2000) 95-102.
\bibitem{SSWW}H. Samtleben, E. Sezgin, R. Wimmer,  JHEP 1112 (2011) 062; JHEP 1303
(2013) 068; H. Samtleben, E. Sezgin, R. Wimmer, L. Wulff, {\it New
superconformal models in six dimensions: Gauge group and
representation structure},  PoS CORFU2011 (2011) 071.
\bibitem{hierar}B. de Wit, H. Samtleben,  Fortsch.Phys. 53 (2005) 442-449.
\bibitem{IB13}I.A. Bandos,  JHEP 1311 (2013) 203.
\bibitem{Z86}B.M. Zupnik,  Sov.J.Nucl.Phys. 44 (1986) 512, Yad.Fiz. 44 (1986)
794-802.
\bibitem{ISZ}E. A. Ivanov, A. V. Smilga and B. M. Zupnik,  Nucl. Phys. B 726
(2005) 131; E.A. Ivanov, A.V. Smilga,  Phys.Lett. B637 (2006)
374-381.
\bibitem{Sok}E. Sokatchev,  Class.Quant.Grav. 5 (1988) 1459-1471; E. Bergshoeff, E. Sezgin, E. Sokatchev,   Class.Quant.Grav. 13 (1996) 2875-2886.
\bibitem{M2D2}S. Mukhi, C. Papageorgakis,  JHEP 05 (2008) 085.
\bibitem{BIaction}E. Bergshoeff, M. Rakowski, E. Sezgin, Phys.Lett. B185 (1987)
371; A.A. Tseytlin, {\it Born-Infeld action, supersymmetry and
string theory}, In *Shifman, M.A. (ed.): The many faces of the
superworld* 417-452, arXiv:hep-th/9908105; E. Bergshoeff, A. Bilal,
M. de Roo, A. Sevrin,  JHEP 0107 (2001) 029.
\bibitem{embed}I. Bandos, K. Lechner, A. Nurmagambetov, P. Pasti, D. Sorokin, M. Tonin, Phys.Rev.Lett. 78 (1997) 4332-4334.
\bibitem{BIK}J. Bagger, A. Galperin, Phys. Lett. B336 (1994) 25, Phys. Rev. D55 (1997) 1091, Phys. Lett. B412 (1997) 296;
S. Bellucci, E. Ivanov, S. Krivonos,
 Phys.Lett. B460 (1999) 348-358; Fortsch.Phys. 48 (2000) 19-24.
\bibitem{ket98}
S.V. Ketov,  Nucl.Phys. B553 (1999) 250-282; M. Rocek, A.A.
Tseytlin,  Phys.Rev. D59 (1999) 106001.
\bibitem{schwarzJ}J.H. Schwarz,  JHEP
1401 (2014) 088.
\bibitem{D4N4}I.L. Buchbinder, S.M. Kuzenko, A.A. Tseytlin,
 Phys.Rev. D62 (2000) 045001;
S.M. Kuzenko,  JHEP 0503 (2005) 008.
\bibitem{gios}A. S. Galperin, E. A. Ivanov, V. I. Ogievetsky and E. S. Sokatchev, {\it Harmonic Superspace},
Cambridge University Press, Cambridge, 2001.
\bibitem{HKcone}P.S. Howe, G. Sierra, P.K. Townsend, Nucl.Phys. B221 (1983) 331;
G. Sierra, P.K. Townsend, Nucl.Phys. B233 (1984) 289; B. de Wit, B. Kleijn, S. Vandoren, Nucl.Phys. B568 (2000) 475-502.
\bibitem{BUCH}I.L. Buchbinder, E.I. Buchbinder, E.A. Ivanov, S.V. Kuzenko, B.A.
Ovrut,
Phys.Lett. B412 (1997) 309-319;  I.L. Buchbinder, E.I. Buchbinder,
S.M. Kuzenko, B.A. Ovrut,
Phys.Lett. B417 (1998) 61-71;
 E.I. Buchbinder, B.A. Ovrut, I.L. Buchbinder, E.A. Ivanov,
S.M. Kuzenko,
Phys.Part.Nucl. 32 (2001) 641-674.
Fiz.Elem.Chast.Atom.Yadra 32 (2001) 1222-1264.
\bibitem{KUZ}
S.M. Kuzenko, I.N. McArthur, Phys.Lett. B513 (2001) 213-222;
Phys.Lett. B506 (2001) 140-146;  JHEP 0305 (2003) 015; S.M. Kuzenko, Phys.Lett.
B600 (2004) 163-170;  Phys.Lett. B644 (2007) 88-93.



\end{thebibliography}
\end{document}